\documentclass[12pt]{article}
\usepackage{graphicx}
\usepackage{epsfig}
\usepackage{color}

\def\lsim{\raise0.3ex\hbox{$<$\kern-0.75em\raise-1.1ex\hbox{$\sim$}}}
\def\gsim{\raise0.3ex\hbox{$>$\kern-0.75em\raise-1.1ex\hbox{$\sim$}}}

\begin {document}
\newcommand{\bt}{\begin{tabular}}
\newcommand{\et}{\end{tabular}}
\newcommand{\bc}{\begin{center}}
\newcommand{\ec}{\end{center}}
\begin{center}

{\bf\large Pseudoscalar meson photoproduction on nucleon target}
\vspace{.5cm}

{\bf G.H. Arakelyan$^1$, C. Merino$^2$ and Yu.M. Shabelski$^{3}$}\\
\vspace{.5cm}

$^1$A.I.Alikhanyan Scientific Laboratory\\
Yerevan Physics Institute\\
Yerevan 0036, Armenia\\
e-mail: argev@mail.yerphi.am\\
\vspace{.1cm}

$^2$Departamento de F\'\i sica de Part\'\i culas, Facultade de F\'\i sica\\
and Instituto Galego de F\'\i sica de Altas Enerx\'\i as (IGFAE)\\
Universidade de Santiago de Compostela\\
15782 Santiago de Compostela\\
Galiza, Spain\\
e-mail: merino@fpaxp1.usc.es \\
\vspace{.1cm}

$^{3}$Petersburg Nuclear Physics Institute\\
NCR Kurchatov Institute\\
Gatchina, St.Petersburg 188350, Russia\\
e-mail: shabelsk@thd.pnpi.spb.ru
\vskip 0.9 truecm

\vspace{1.2cm}
\vskip 0.5 truecm

{\bf Abstract}
\end{center}


We consider the photoproduction of secondary mesons in the framework of
the Quark-Gluon String model. At relatively low energies, not only cylindrical, but also planar
diagrams have to be accounted for. To estimate the significant contribution of planar diagrams
in $\gamma p$ collisions at rather low energies, we have used the expression obtained from the
corresponding phenomenological expression for $\pi p$ collisions. The results obtained by the model
are compared to the existing SLAC experimental data. The model predictions for light meson production
at HERMES energies are also presented.  

\newpage

\section{Introduction}

The physical photon state is approximately represented by the superposition of a bare photon and of virtual 
hadronic states having the same quantum numbers as the photon. The bare photon may interact 
in direct processes. This direct contribution is determined by the lowest order in perturbative QED. 
The interactions of the hadronic components of the photon (resolved photons) at
high enough energies are described within the framework of theoretical models based on Reggeon 
and Pomeron exchanges, like the Dual-Parton model (DPM), see~\cite{capella} for a review, and
the Quark-Gluon String model (QGSM)~\cite{K20,KTM}. 

The QGSM is based on the Dual Topological Unitarization (DTU) and on the large 1/N expansion of 
non-perturbative QCD, and it has been very successful in describing many features of the
inclusive hadroproduction of secondaries, both at high~\cite{KTM}-\cite{Sh}, and
at comparatively low~\cite{volk,kaidlow,amelin} energies. 

In the case of photoproduction processes, QGSM was used in \cite{lugovoi,aryer} at energies 
higher then 100 GeV. At these high energies the accounting for only pomeron (cylindrical) diagrams 
leads to a reasonable agreement. However, at low energies one can nott neglect the contribution
of planar diagrams~\cite{volk,kaidlow,amelin}. The inclusion of palnar diagrams into the analysis of
pp scattering should result in the better description of a wider range of experimental data.  

Most interesting it is the comparison of secondary production by photon and hadron beams 
in their fragmentation regions. The photoproduction ($Q^2$ = 0) points are the boundary
($Q^2\rightarrow$0) for the electroproduction processes~\cite{badelek}.

In the present paper we apply QGSM to the low energy photoproduction of pions and kaons by taking 
into accaount, contrary to~\cite{lugovoi,aryer}, both cylindrical and planar diagrams. 
We describe the SLAC experimental data~\cite{pi0100,abe20} on inclusive  $x_F$ spectra integrated over 
$p_T$. The predictions for forthcoming quasireal photoproduction data at HERMES and JLAB energies 
are also presented. 

\section{Inclusive spectra of secondary hadrons in photoproduction processes}

Let us now consider the photoproduction processes in the QGSM.
The resolved photon can be consider as 
\begin{equation}
\label{vdm2}
|\gamma> = \frac{1}{6}(4|\overline{u}u> + |\overline{d}d> + |\overline{s}s>) \; ,
\end{equation}
in agreement with standard Vector Dominance Model (VDM) expansion~\cite{bauer}, and so
the quarks in Eq.~\ref{vdm2} can be considered as the valence quarks of a meson. 
This resolved photon state also has sea quarks and gluons, as usual hadrons.  

We can calculate the photoproduction cross section like for a meson-nucleon process\footnote
{The absolute values of secondary photoproducton cross section are $\alpha_{em}$ times
those of meson-nucleon cross section, while the corresponding multiplicities of secondaries
$\frac{dn}{dy}$ are of the same order.} by summing, according to Eq.~\ref{vdm2},
the contribution of $q\overline{q}$ pair collisions with target nucleon, and by taking into
account the corresponding coefficients.

At low energies, the inclusive cross section in QGSM consists of two terms, described by two 
different types of diagrams: cylindrical and planar. 
\begin{figure}[htb]
\label{onep}
\begin{center}
\vskip -12.5cm
\includegraphics[width=1.25\hsize]{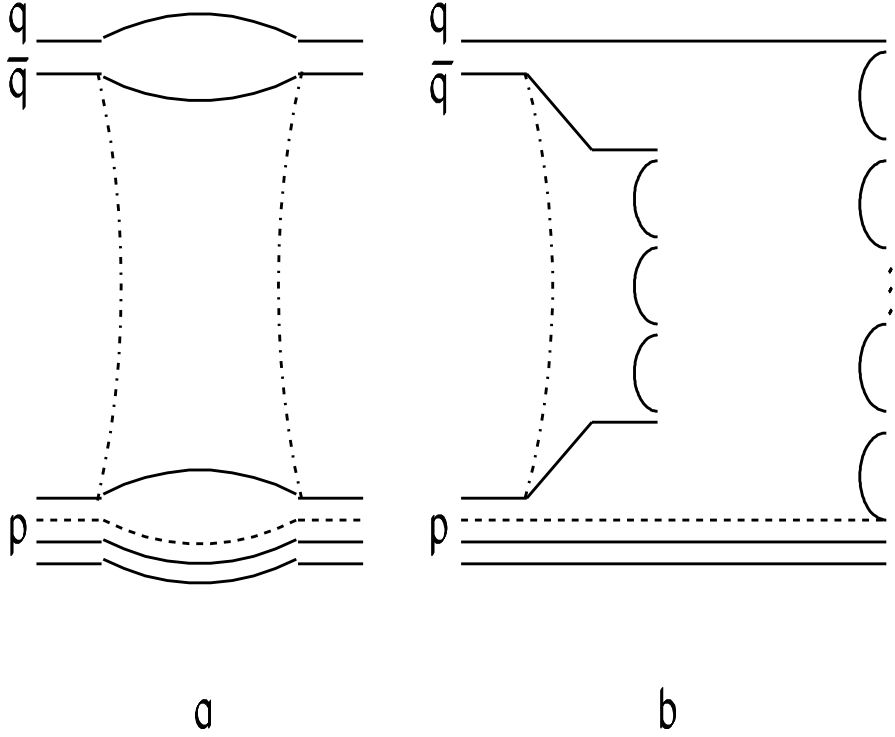}
\vskip -1.5cm
\caption{\footnotesize
(a) Cylindrical diagram corresponding to the one-Pomeron exchange contribution to 
elastic $q\overline{q}$ scattering, and (b) the cut of this diagram which determines the 
contribution to the inelastic $q\overline{q}$ cross section. Quarks are shown by 
solid curves, while SJ is shown by dashed curves.}
\end{center}
\end{figure}
\begin{equation}
\label{spectrt}
\frac{dn}{dy}\ = \frac{1}{\sigma_{inel}}\frac{d\sigma}{dy}\ = \frac{dn^{cyl}}{dy} + \frac{dn^{pl}}{dy}
\end{equation}
In QGSM high energy interactions are considered as takin place via the exchange of one or 
several Pomerons, and all elastic and inelastic processes result from cutting through 
or between pomerons~\cite{AGK} (see Fig.1). 
Each Pomeron corresponds to a cylindrical diagram (see Fig.~1a), and thus, when cutting 
a Pomeron, two showers of secondaries are produced, as it is shown in Fig.~1b. 

For $\gamma p$ interaction, that following Eq.~\ref{vdm2} is similar to 
$\pi p$ collisions, the inclusive spectrum of a secondary hadron $h$ produced from a 
$q \overline{q}$ pair has the form~\cite{KaPi,Sh}:
\begin{equation}
\label{spectr}
\frac{dn^{cyl}}{dy}\ =\frac{x_E}{\sigma_{inel}} \frac{d\sigma^{cyl}}{dx_F}\ =\ \sum_{n=1}^\infty
w_n\phi_n^h (x)\ ,
\end{equation}
where $x_{F}=2p_{\|}/\sqrt{s}$ is the Feynman variable, $x_{E}=2E/\sqrt{s}$, 
the functions $\phi_{n}^{h}(x)$ determine the contribution of diagrams with $n$ cut 
Pomerons, and $w_n$ is the relative weight of these diagrams. Thus,
\begin{eqnarray}
\label{spectr1}
\phi_{q\overline{q}p}^h(x) &=& f_{\overline{q}}^{h}(x_+,n)\cdot f_q^h(x_-,n) +
f_q^h(x_+,n)\cdot f_{qq}^h(x_-,n)
\nonumber\\
&+&2(n-1)f_{sea}^h(x_+,n)\cdot f_{sea}^h(x_-,n)\ ,
\\
x_{\pm} &=& \frac12\left[\sqrt{4m_T^2/s+x^2}\ \pm x\right] ,
\end{eqnarray}
where $f_{qq}$, $f_q$, and $f_{sea}$ correspond to the contributions of diquarks, valence quarks 
and sea quarks, respectively.
These functions $f_{qq}$, $f_q$, and $f_{sea}$ are determined by the convolution of the diquark 
and quark distribution functions, $u(x,n)$, with the fragmentation functions, $G^h(z)$, to hadron $h$, 
e.g.
\begin{equation}
f_i^h(x_+,n)\ =\ \int\limits_{x_+}^1u_i(x_1,n)G_i^h(x_+/x_1) dx_1\; , 
\end{equation}
where $i=q, \overline{q}, qq$-diquarks, and sea quarks.

The fragmentatiion functions $G_i^{\pi}(z)$ have the same value, $a^\pi$ at $z \rightarrow 0$ for  
all $i$. Similarly, $G_i^K(z)$  have also the same value, $a^K$ at $z \rightarrow 0$ for all $i$.
The numerical values of these parameters are~\cite{ampsh1}:
\begin{equation}
\centering
a^\pi = 0.68,\  a^K=0.26\; .
\end{equation}
The diquark and quark distribution functions, which are normalized to unity, 
as well as the fragmentation functions, are determined from Regge intercepts~\cite{kaidff}. 
The analytical expressions of these functions for proton are presented in~\cite{KaPi,Sh}. 
The distribution functions for quarks and antiquarks in a photon were obtained by using the 
simplest interpolation of Regge limits at $u_i(x\rightarrow 0)$ and $u_i(x\rightarrow 1)$, 
following~\cite{KaPi,Sh}. In the sum of all cylindrical diagrams we have used the weights given
in Eq.~1. 

At low energies the contribution of planar diagrams becomes significant. In particular, planar 
diagrams lead to the difference from $\sigma^{\pi^- p}_{tot}$ (Fig.~2a)
to $\sigma^{\pi^+ p}_{tot}$(Fig.~2b) total cross sections.

Since the proton contains two $u$ quarks and one $d$ quark, there are two planar diagrams contributing
to $\sigma_{tot}^{\pi^- p}$ (Fig.~2a), and only one contributing to $\sigma_{tot}^{\pi^+ p}$
(Fig.~2b). By neglecting the difference in $u$$\overline{u}$ and $d$$\overline{d}$
annihilation, we can consider as equal the contribution by every planar diagram. If
we denote this contribution by each planar digram as $\sigma^{\pi p}_{pl}$, the
contribution of diagrams in Fig.~2a to the $\sigma_{tot}^{\pi^- p}$ is
2$\sigma^{\pi p}_{pl}$,
while the contribution of Fig.~2b to the total
$\sigma_{tot}^{\pi^+ p}$ cross section is $\sigma^{\pi p}_{pl}$. Thus,
\begin{equation}
\Delta \sigma(\pi^\mp p) = \sigma_{tot}(\pi^- p) -  \sigma_{tot}(\pi^+ p) = \sigma_{pl}^{\pi p} \; ,
\end{equation}
the cylindrical contributions cancelling each other off into the difference. 
\begin{figure}[ht]
\label{plsigt}
\center
\vskip -17.cm
\includegraphics[width=1.5\hsize]{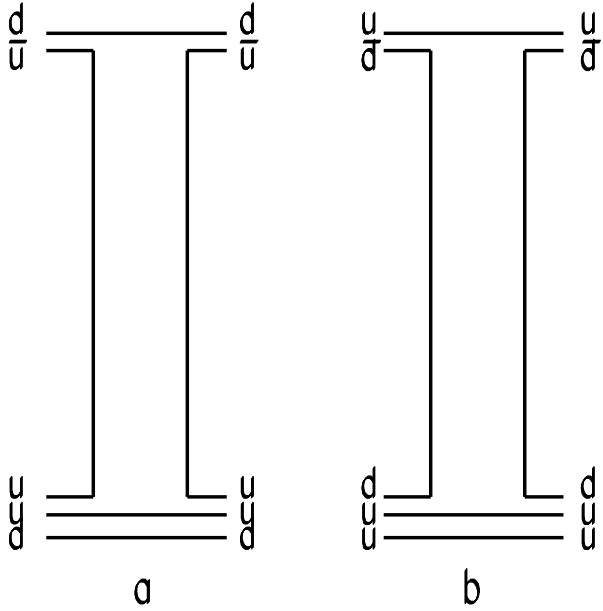}
\vskip -4.cm
\caption{Planar diagrams for the (a) $\pi^- p$ and (b) $\pi^+ p$ 
elastic scattering amplitudes.}
\end{figure}
\begin{figure}[htb]
\label{planar}
\begin{center}
\vskip -17.cm
\includegraphics[width=1.5\hsize]{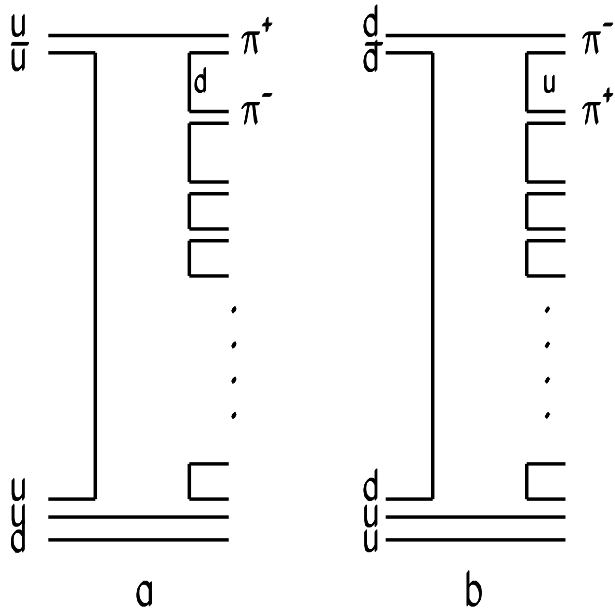}
\vspace{-4.cm}
\caption{Planar diagrams describing secondary meson $M$ production (a) by $u$ and 
(b) by $d$ valence quarks from photon.}
\end{center}
\end{figure}
One can find $\overline{u}$ quark in the photon with probability 
$\frac{4}{6}$ (see Eq.~1), and so, simply from comparison of the number of diagrams of figs. 2a and 3a 
at the same energies, there is a contribution to planar photoproduction cross section 
from the diagram Fig.~3a (strange quarks do not contribute to planar photoproduction) equal to
\begin{equation}
\frac{\sigma^{\gamma p(3a)}_{pl}}{\sigma_{inel}^{\gamma p}} = \frac{4}{3}
\frac{\sigma_{pl}^{\pi p}}{\sigma_{inel}^{\pi p}} \; ,
\end{equation}
and, in a similar way, the contribution to planar photoproduction cross section from the diagram Fig.~3b 
is
\begin{equation}
\frac{\sigma^{\gamma p(3b)}_{pl}}{\sigma_{inel}^{\gamma p}} = \frac{1}{6}
\frac{\sigma_{pl}^{\pi p}}{\sigma_{inel}^{\pi p}}\; , 
\end{equation}
with $\sigma^{\gamma p}_{tot} \cong \sigma_{inel}^{\gamma p}$ in both cases.
Thus, the resulting contribution from planar photoprouction coming from diagrams in figs.~3a
and 3b to the inclusive cross section is determined by using similar formula to those
in~\cite{K20,kaidlow}:
\begin{eqnarray}
\label{plan}
\centering
\frac{dn^{\gamma p}_{pl}}{dy}\ &=& \frac{\sigma^{\gamma p}_{pl}}{\sigma_{inel}^{\gamma p}}
[\frac{4}{3}G_u^h(x_+)G_{ud}^h(x_-)+ \frac{1}{6}G_d^h(x_+)G_{uu}^h(x_-)] \\
&=& \frac{\Delta \sigma(\pi^\mp p)}{\sigma_{inel}^{\pi p}}[\frac{4}{3}G_u^h(x_+)G_{ud}^h(x_-)+
\frac{1}{6}G_d^h(x_+)G_{uu}^h(x_-)]
\end{eqnarray}

The parametrisation of experimental data on $\Delta \sigma({\pi^\mp p})$ at 
$\sqrt{s} \geq$ 5GeV exists~\cite{pdg}:
\begin{equation}
\Delta \sigma({\pi^\mp p}) = 2.161(\frac{s}{s_M})^{-0.544}{\rm mb}\; , 
\end{equation}
where $s_M = (\mu + m_p + 2.177)^2$, $\mu$ is the pion mass, and $m_p$ is the proton mass.
At energies $\sqrt{s}\approx 5 GeV$, where the contribution of planar diagrams is not
negligible, $\sigma_{inel}^{\pi p} \approx$ 26 mb. 



For $K$-mesons photoproduction, the planar diagrams exist only for leading $K^+$ production, 
since only the u-quark in the photon can create the planar diagram in the collision with the
target proton. On the other hand, the $K^-$-meson can be produced in a planar diagram as a slower
nonleading particle.

\section{Results of calculations}

In this section, we present the results of the QGSM calculation for pseudoscalar meson
photoproduction on a proton target. Thus, in Fig.~4 we compare the QGSM results with the
experimental data on $\pi^0$ photoproduction obtained by OMEGA collaboration~\cite{pi0100}.
In this experiment the cross sections were measured at photon energies of 110$-$135 GeV,
85$-$110 GeV, and 50$-$85 GeV. The theoretical curves in Fig. 4 have been calculated at
photon energies 120 GeV (dashed line), 100 GeV (full line), and 70GeV (dashed-dotted line).
As we can see, the theoretical curves obtained at these three energies practically coincide,
in correspondence to the scaling shown by the experimental data.
At these energies the contribution of planar diagrams is small, and the present calculations
are close to the results by~\cite{aryer}, where the only contribution from cylindrical diagrams
was taken into account. 

In Fig. 5 we present the comparison of QGSM calculations for the $x_F$ dependence of the $K^0_S$
inclusive cross section $F(x_F)=(1/\sigma_{tot})d\sigma/dx_F$, integrated over $p_T^2$, 
to the experimental data by~\cite{abe20} at $E_{\gamma}$= 20 GeV. 
The dashed line corresponds to the contribution from only cylindrical diagrams, while
the full curve represents the sum of contributions from both cylindrical and planar diagrams.
As we can see, though the contribution of the planar diagrams is not large, its inclusion
leads to a better agreement with the experimental data. However, one has to note that the
theoretical result is proportional to $(a^K)^2$ (Eq.~7), and the value of this parameter
is mainly known from high energy pp collisions, so its accuracy is estimated not to be
better then 10\%~\cite{ampsh1}.

\begin{figure}[htb]
\label{pi0}
\centering
\vskip -10.cm
\includegraphics[width=1.25\hsize]{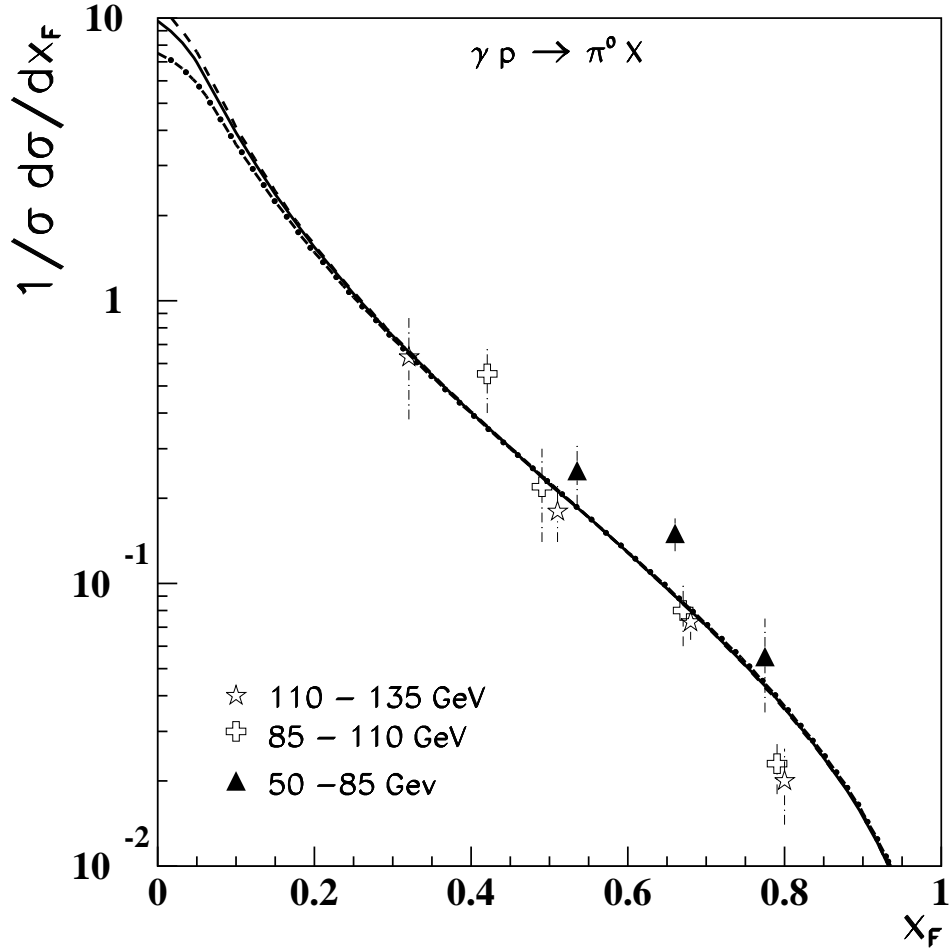}
\vskip -1.cm
\caption{QGSM calculation of the $x_F$ dependence of the $\pi^0$ photoproduction cross section 
integrated over $p_T^2$, compared to the experimental data~\cite{pi0100}.
The full line corresponds to calculations at a photon energy of 100 GeV.} 
\end{figure}
\begin{figure}[htb]
\label{k0}
\centering
\vskip -6.5cm
\includegraphics[width=1.25\hsize]{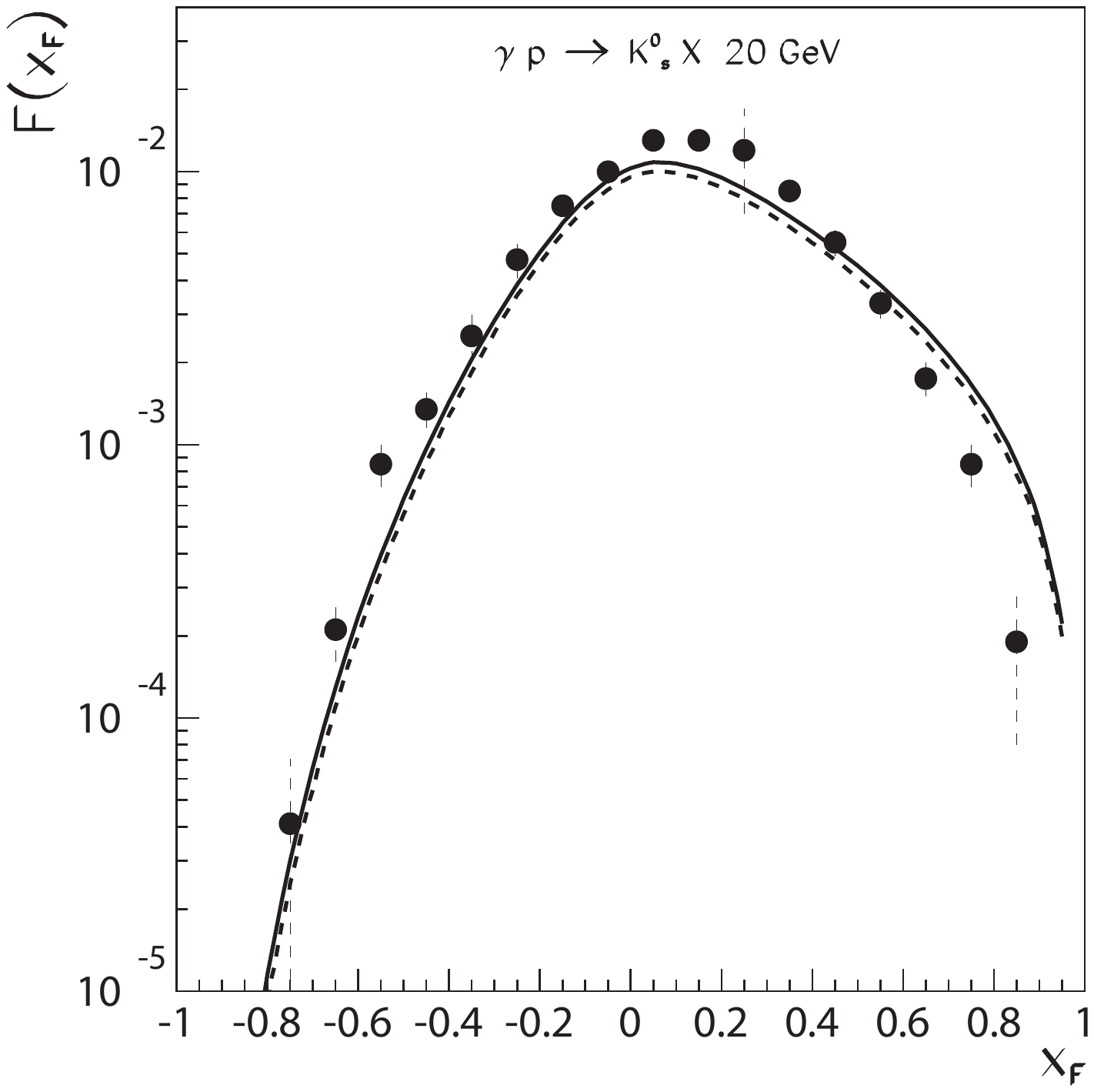}
\vskip -5.5cm
\caption{
QGSM predictions for the $x_F$ dependence of $K^0_S$ photoproduction cross section 
integrated over $p_T^2$, and compared to the experimental data~\cite{abe20}.
The full line corresponds to the calculations for a photon energy of 20 GeV,
and the dashed line to the corresponding contribution from only cylindrical diagrams.} 
\end{figure}
In Fig. 6 we show the model predictions for the inclusive density of $\pi^\pm$ and $K^\pm$ mesons at 
the energy of HERMES Collaboration $E_{\gamma}$= 17 GeV. The full curves correspond to the inclusive 
spectra of $\pi^+$ and $K^+$, while dashed lines represent $\pi^-$ and $K^-$ meson photoproduction. 
We show the summed contribution of both cylindrical and planar diagrams.
\begin{figure}[htb]
\label{pi20}
\centering
\vskip -6.5cm
\includegraphics[width=1.25\hsize]{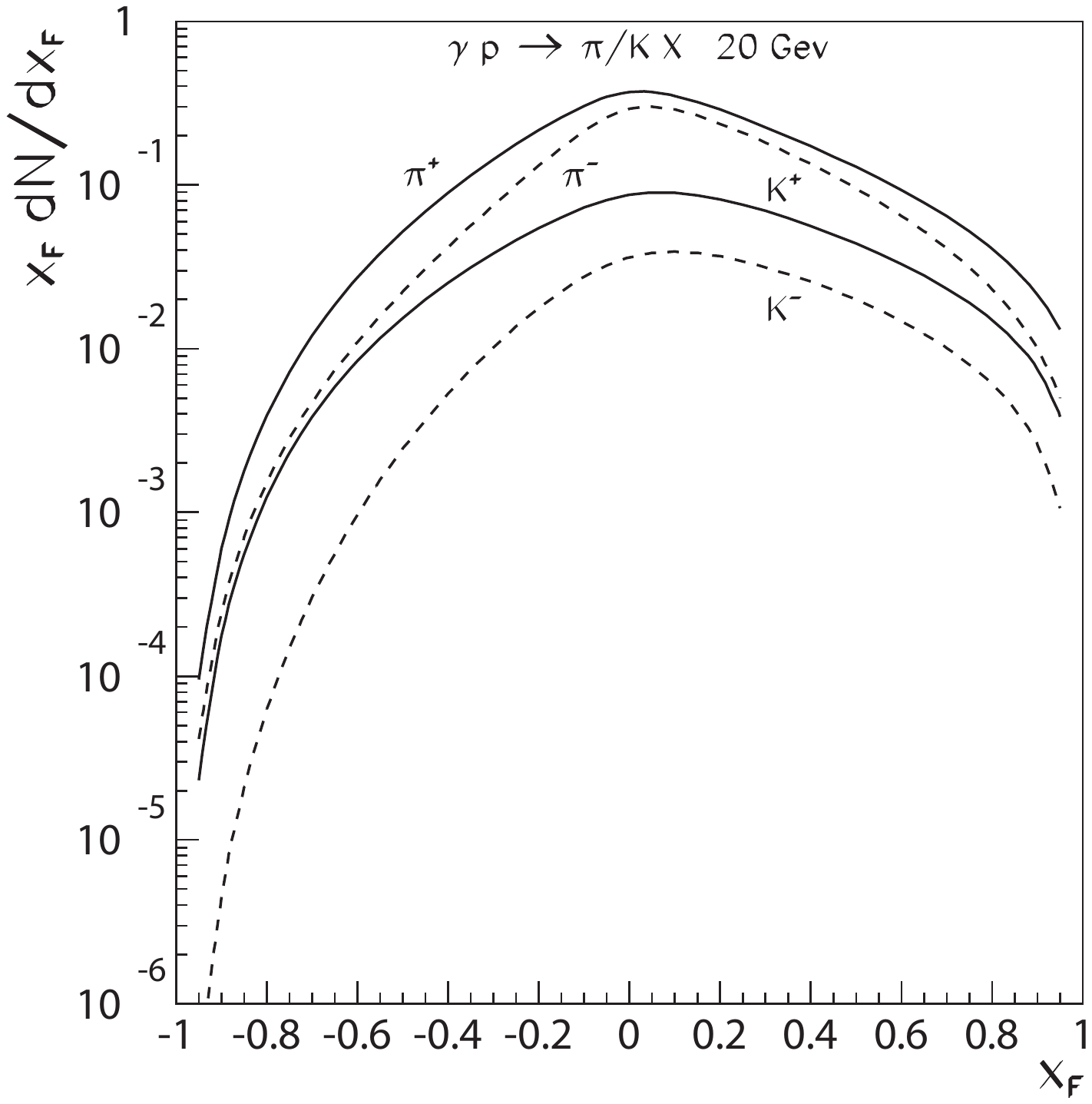}
\vskip -5.5cm
\caption{The QGSM prediction for the $x_F$ dependence of the invariant
cross section $F(x_F)=1/\sigma_{tot}d\sigma/dx_F$ integrated over $p_T^2$ 
spectra of $\pi^+$ (full) and $\pi^-$ (dashed), upper lines, and
of $K^+$ (full) and $K^-$ (dashed), lower lines,
photoproduction at $E_{\gamma}$= 17 GeV.} 
\end{figure} 
\begin{figure}[htb]
\label{rpi}
\centering
\vskip -10.5cm
\includegraphics[width=1.25\hsize]{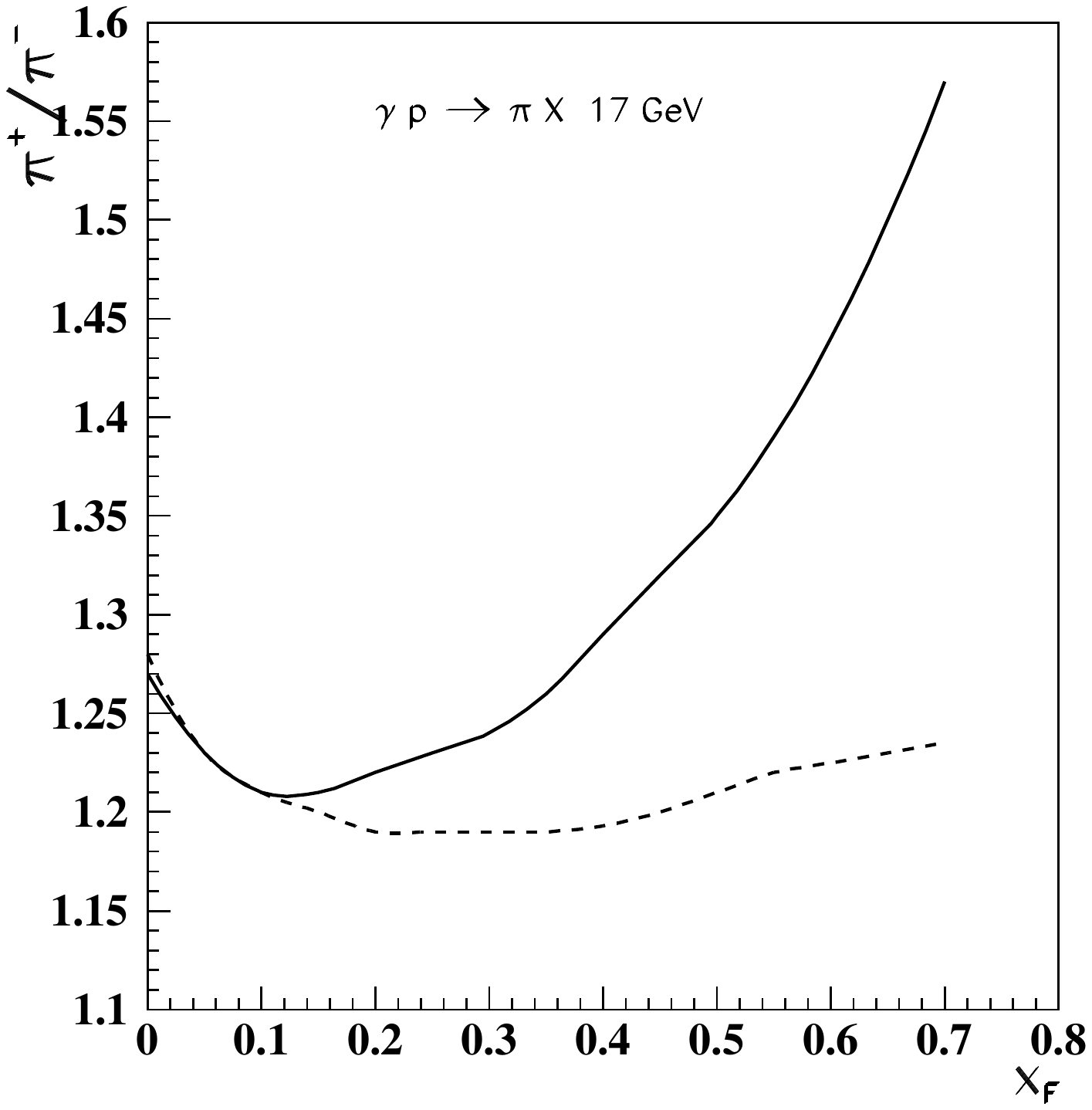}
\vskip -1.5cm
\caption
{The QGSM prediction for the $x_F$ dependence of the ratio of yields of charged 
$\pi$ mesons at $E_{\gamma}$= 17 GeV (full line). The dashed line shows the contribution
from only the cylindical diagram.}
\end{figure}
\newpage
The prediction for the ratio of yields of $\pi^+/\pi^-$-mesons at $E_{\gamma}$= 17 GeV 
is shown in Fig. 7 by full line. The cylindrical contribution is shown by dashed line. 
One can see that the planar diagram contribution changes the ratio by 25\% in the
forward hemisphere at large $x_F$.
%

In Fig. 8 the ratio of yields of $K^+/K^-$-mesons at $E_{\gamma}$= 17 GeV is shown.
Only valence $u$-quark contributes for leading $K^+$-meson production in planar diagram. 
\begin{figure}[htb]
\label{rpi}
\centering
\vskip -10.5cm
\includegraphics[width=1.25\hsize]{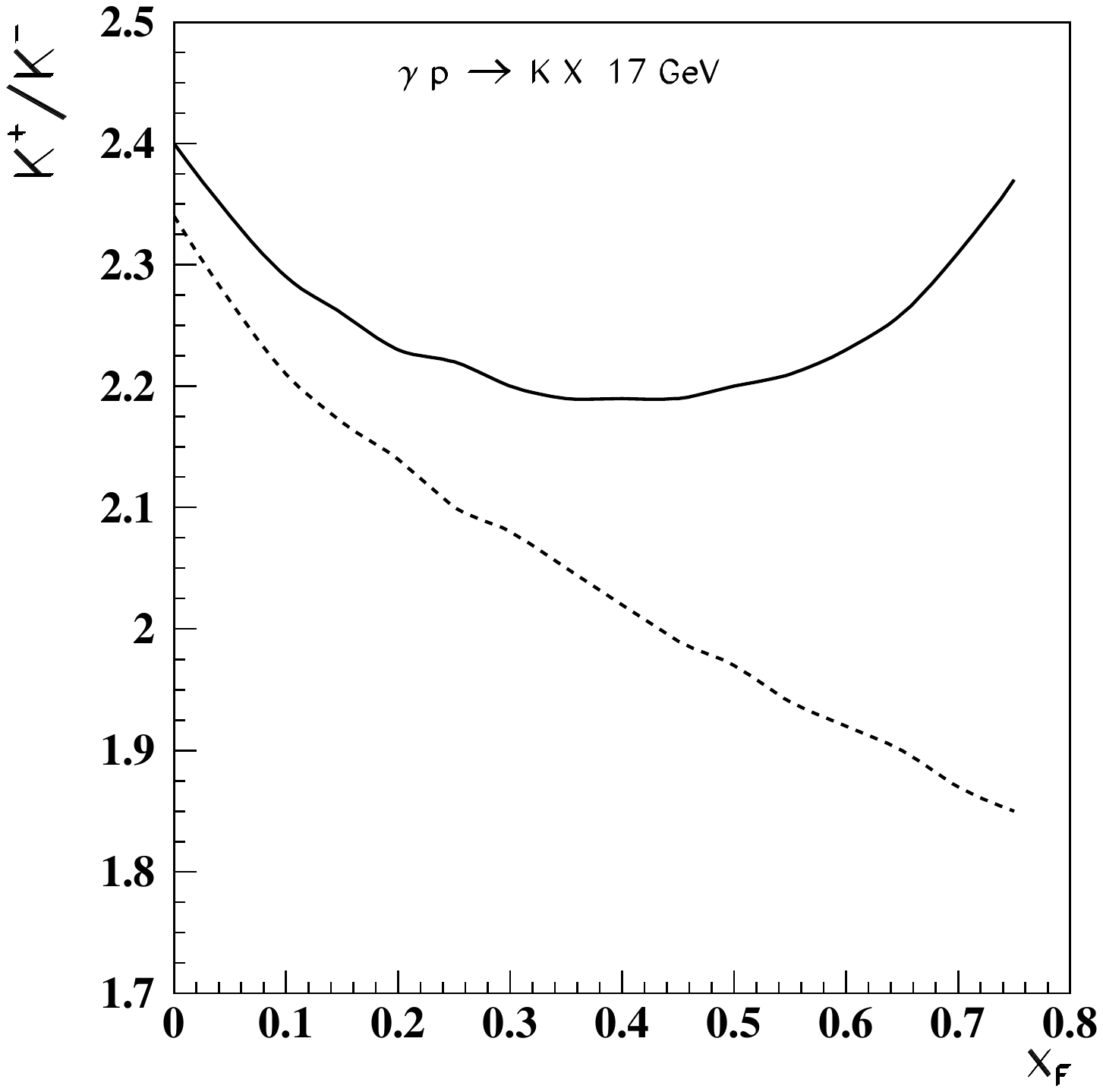}
\vskip -1.cm
\caption{The QGSM prediction for the $x_F$ dependence of the ratio of yields of charged 
$K$-mesons photoproduction at $E_{\gamma}$= 17 GeV.}
\end{figure}

\section{Conclusion}
We consider a modified QGSM approach for the description of pseudoscalar ($\pi$, $K$)
mesons photoproduction on nucleons at relatively low energies. 
This approach gives reasonable agreement to the experimental data on the $x_F$ dependence
for $\pi ^0$ cross sections at $E_{\gamma}$=100 GeV, and for $K^0_S$ at $E_{\gamma}$=20 GeV, by 
taking into account the planar diagrams contribution that becomes significant at low energies.
We also present the model predictions for charged pions and kaons cross sections and
for the yields ratios at $E_{\gamma}$=17 Gev (Hermes Collaboration energies). 

The comparison of the model results with experimental data allows the estimation of the contribution
of the planar diagrams to the particle photoproduction processes.
Detailed comparison of the theoretical predictions to experimental data at low energies makes possible
to refine the values of the parameters of the model, and, in this way, it improves the reliability
of the calculated yields of secondary particles to describe possible future HERMES data
of quasireal photoproduction processes.  

\newpage

{\bf Acknowledgements}

We are grateful to N.Z. Akopov, G.M. Elbakyan, and P.E. Volkovitskii for useful discussions. 
This paper was supported by the State Committee of Science of Republic of Armenia, Grant-13-1C015, 
and by Ministerio de Econom{\'i}a y Competitividad of Spain (FPA2011$-$22776),
the Spanish Consolider-Ingenio 2010 Programme CPAN (CSD2007-00042), and
Xunta de Galicia, Spain (2011/PC043).


\newpage

\begin{thebibliography}{**}


\bibitem{capella} A. Capella, U. Sukhatme, C.I. Tan, and J. Tran Thanh Van, 
Phys. Rep. {\bf 236}, 225 (1994).   

\bibitem{K20}  A.B. Kaidalov, Phys. Atom. Nucl., {\bf 66}, (2003), 781.

\bibitem{KTM} A.B. Kaidalov and K.A. Ter-Martirosyan, Yad. Fiz. {\bf 39}, 1545 (1984); 
Yad. Fiz. {\bf 40}, 211 (1984).

\bibitem{acks} G.H. Arakelyan, A. Capella, A.B.~Kaidalov, and Yu.M.~Shabelski, Eur. Phys. J.
{\bf C26}, 81 (2002).

\bibitem{KaPi} A.B. Kaidalov, O.I. Piskunova, Yad. Fiz. {\bf 41}, 1278 (1985).

\bibitem{Sh} Yu.M. Shabelski, Yad. Fiz. {\bf 44}, 186 (1986).

\bibitem{volk} P.E. Volkovitski, Yad. Fiz. {\bf 44}, 729 (1989).

\bibitem{kaidlow} A.B. Kaidalov {\it et al.}, Yad. Fiz. {\bf 49}, 781 (1989); 
Yad. Fiz. {\bf 40}, 211 (1984).

\bibitem{amelin} N.S. Amelin {\it et al.}, Yad. Fiz. {\bf 51}, 941 (1990); 
Yad. Fiz. {\bf 40}, 211 (1984).

\bibitem{lugovoi} V.V. Luigovoi, S.Yu. Sivoklokov, Yu.M. Shabelskii. Phys. Atom. Nucl.
{\bf 58}, 67 (1995); Yad. Fiz. {\bf 58} (1995) 72.

\bibitem{aryer} G.G. Arakelyan and Sh.S. Eremian, Phys. Atom. Nucl. {\bf 58},
1241 (1995); Yad. Fiz. {\bf 58}, 132 (1995).

\bibitem{badelek} B. Badelek and J. Kwiecinski, Rev. Mod. Phys. {\bf 68} 445 (1996).

\bibitem{pi0100} R.J. Apsimon {\it et al.}, Z. Phys. {\bf C52}, 397 (1991). 

\bibitem{abe20} K. Abe {\it et al.}, Phys. Rev. {\bf D32}, 2869 (1985).

\bibitem{bauer} T.H. Bauer {\it et al.}, Rev. Mod. Phys. {\bf 50}, 261 (1978).

\bibitem{ampsh1} G.H. Arakelyan, C. Merino, C. Pajares, and Yu.M.~Shabelski, Eur. Phys. J.
{\bf C54}, 577 (2008); arXiv:0805.2248 [hep-ph].

\bibitem{pdg} Particle Data Group, Chin. Phys. {\bf C38}, 090001 (2014).

\bibitem{AGK} V.A. Abramovski, V.N. Gribov, and O.V. Kancheli, Yad. Fiz. {\bf 18}, 595 (1973). 

\bibitem{kaidff} A.B. Kaidalov. Yad. Fiz. {\bf 45}, 1432 (1987).

\bibitem{gandsm2} J. Gandsman {\it et al.}, Phys. Rev. {\bf D10}, (1974) 1562.

\end{thebibliography}
\end{document}